\documentclass{elsarticle}             

\usepackage{amsmath}
\usepackage{amssymb}
\usepackage{bm}
\usepackage{graphicx}

\def\be{\begin{equation}}
\def\ee{\end{equation}}
\def\bea{\begin{eqnarray}}
\def\eea{\end{eqnarray}}

\begin{document}

\begin{frontmatter}

\title{Extinction by the long  dielectric needles
} 
\author[Cherkas1]{S. L. Cherkas}\ead{cherkas@inp.minsk.by}    
\author[Cherkas2]{N. L. Cherkas}
\address[Cherkas1]{Institute for Nuclear Problems, Belarus State University, Bobruiskaya 11, Minsk,
220050 Belarus}  
\address[Cherkas2]{Military Academy of the Republic of Belarus,
Nezavisimosti Av. 220,Minsk,  220057 Belarus
}             


\begin{abstract}                          
Electromagnetic wave extinction by the very long but finite
dielectric needle is compared with that by the infinite dielectric
cylinder for an oblique incidence of the electromagnetic wave. It
is shown that the renormalized Hankel functions without the
logarithmic terms should be used for the calculation of the
extinction per unit length of the infinite dielectric cylinder to
apply it for extinction calculations  by the  finite dielectric
cylinder.
\end{abstract}

\end{frontmatter}

\section{Introduction}

The problem of the scattering of the electromagnetic wave  by the
finite dielectric cylinder often arises in the optics and
radiophysics. Last time the problem acquires a new aspect in a
light of the optical properties of the synthetic nanomediums which
contain the long dielectric cylinders or pores as the structure
elements \cite{nano}. In particular, the matrix of anodized
aluminum oxide \cite{nano1} allows to obtain the arrays of
nanorods by filling pores of the matrix by the metals or other
substances \cite{moon}. The typical problem is a falling of the
electromagnetic wave on a layer (or a number of layers) consisting
of the parallel dielectric cylinders (nanowires or nanorods) or
cylindrical pores. Usually the cylinders are perpendicular to the
surface of the layer, and in the general case the electromagnetic
wave has an oblique incidence relative to the  axis of the
cylinders. The typical width of a layer (i.e. length of the
cylinder) is of the hundreds of micrometers and the cylinder
radius is of order tens nanometers. Thus, the aspect ration is
$L/(2 R)\sim 10^4$, where $L$ and $R$ are the length and radius of
the cylinder respectively. From the other hand, a detector is
usually situated at a macroscopical distance from the layer,that
is, an every cylinder should be considered as three dimensional
object in the scattering problem despite of the large aspect
ratio. There is no an analytic solution of the scattering problem
by the finite dielectric cylinder. Moreover, the numerical methods
\cite{num1,num2,num3} are complicated when the aspect ratio is
large. From the other hand, the scattering amplitude by the
infinite cylinder is expressed analytically in a closed form
\cite{hulst, boren}.

\begin{figure}[th]
\hspace{-0. cm}
\includegraphics[width=9.5 cm]{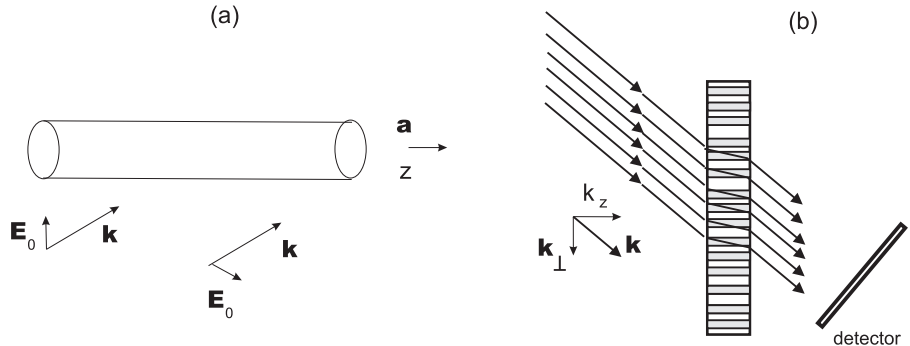}
\caption{\label{fig1} Geometry of the electromagnetic wave
scattering by a single dielectric cylinder (a), wave transmission
through a layer consisting of the cylinders or cylindrical pores
(b).}
\end{figure}

In the monograph \cite{boren} one could read that the formulas for
extinction by the infinite cylinder could be applied for the
finite one when the boundary effects are negligible. But almost in
all the physical situations the boundary effects are considerable!
The boundary effects are negligible when an observer is situated
at the distance much less then the length of the cylinder, where
the field has an asymptotic $1/\sqrt{r}$, however, generally a
finite cylinder has far field falling as $1/r$ in all direction of
the space. Thus, we principally could not avoid the boundary
effects if we are talking about the scattering  on the three
dimensional body.

In Ref. \cite{hulst, wang} on basis of the Huygens's principle
applied at the intermediate zone, where the cylindrical wave
transforms into the spherical one, it was argued that under the
normal incidence of a wave  the extinction cross section by the
finite cylinder per unit length equals to that by infinite
cylinder. However, the Huygens's principle itself is an
approximation.

From the other hand, there exist the approximate methods namely
generalized Rayleigh-Gans (GRG) approximation \cite{shiff0,shiff}
which extends Rayleigh-Gans approximation to the case of the
scatterers  having one large dimension compared to the others. In
the GRG approximation the finite cylinder is considered as  a
three dimensional object. It seems instructive to compare the
extinction in the GRG approximation with that given by the
infinite cylinder. That is done in the present paper. Then we
suggest a new method how to apply the extinction by the infinite
cylinder to finite one.

Certainly there exists the range of the angles where  the finite
cylinder could not modelled by the infinite one. If one looks
along a direction close to the axis of the cylinder it looks like
a circle, while the infinite cylinder formally always looks as an
extended object.

\section{GRG approximation}

Let us remind GRG approximation \cite{shiff0,shiff} because it
will be test bed for comparing with the results for the infinite
cylinders.

An electromagnetic field scattered by the dielectric body is
expressed with the help of  diadic Greene function as \cite{jack}
\be \bm E(\bm
r)=\frac{1}{4\pi}(\varepsilon-1)(\bm\nabla\otimes\bm\nabla+k^2)\int
\frac{\exp(i k|\bm r-\bm r^\prime|)}{|\bm r-\bm r^\prime|}\bm
E(r^\prime)d^3\bm r^\prime+\bm E_0e^{i\bm k\bm r}. \label{first}
\ee At large $r$ one has \bea (\bm \nabla\otimes\bm
\nabla+k^2)\frac{e^{i k|\bm r-\bm r^\prime|}}{|\bm r-\bm
r^\prime|}\bm E(\bm r^\prime)\approx-k^2\frac{e^{i k|\bm r-\bm
r^\prime|}}{|\bm r-\bm r^\prime|}\bm u\times(\bm u\times \bm E(\bm
r^\prime))\nonumber\\\approx -\frac{e^{ik r-i\bm k^\prime \bm
r^\prime}}{r}\bm k^\prime\times(\bm k^\prime\times \bm E(\bm
r^\prime)),\label{appr} \eea where $\bm u=\frac{\bm r-\bm
r^\prime}{|\bm r-\bm r^\prime|}$, $\bm k^\prime=k\frac{\bm r}{r}$.
In the first approximate equality of the Eq. \eqref{appr} the
terms falling faster then the $\frac{1}{|\bm r-\bm r^\prime|}$ are
omitted and in the second approximate equality we set $\bm r-\bm
r^\prime\rightarrow \bm r$ everywhere except for $\exp(i k|\bm
r-\bm r^\prime|)$ which is approximated as $\exp(i k|\bm r-\bm
r^\prime|)\approx \exp(ikr-i\bm k^\prime\bm r^\prime)$.

The next step is to approximate the field $E(\bm r^\prime)$ inside
the scatterer. Let us denote the unit vector $\bm a$ along the
axis of the cylinder as it it shown in Fig.1 (a). The
approximation \cite{shiff0,shiff} approachs the field inside the
cylinder as uniform in the cross section of the cylinder and
depending only on the longitudinal coordinate $z$. This uniform
field is applied equal to the static field as though
 this cylinder would be placed in the external electric field $\bm E_0$.
  That is $\bm E(\bm r^\prime)=\left(\frac{2}{\varepsilon +1}(\bm
E_0-\bm a(\bm E_0 \bm a))+\bm a(\bm E_0 \bm a)\right)e^{i(\bm k
\bm a)(\bm r^\prime\bm a)}$. The integral in the Eq. \eqref{first}
could be calculated and the asymptotic of the scattered field
takes the form
\bea
\bm E(\bm r)\approx-\frac{e^{i k r}}{r}\bm \left(\bm
k^\prime\times\left(\bm k^\prime\times \left(\frac{2}{\varepsilon
+1}(\bm E_0-\bm a(\bm E_0 \bm a))+\bm a(\bm E_0 \bm
a)\right)\right)\right)\nonumber\\\frac{(\varepsilon -1)L
R^2}{2}\,\frac{\sin\left((\bm k-\bm k^\prime)\bm a/2\right)}{(\bm
k-\bm k^\prime)\bm a}.~~~~~
\label{ref}
\eea

The formula \eqref{ref} is identical to that of Ref.
\cite{shiff0}\footnote{Ref. \cite{shiff} contains misprint in the
analogous formula.}  but it is written purely in the vector
form.The differential cross section is defined as \be
\frac{d\sigma}{d\Omega}=r^2\frac{\bm E(\bm r)\bm E^*(\bm
r)}{E_0^2}, \ee where $d\Omega$ is a cone around $\bm k^\prime$.

It is possible to obtain the  total cross section in the
analytical form \cite{shiff0, shiff}. Let us write the cross
section for two different cases: an ordinary wave, when the $\bm
E_0$ is perpendicular to the axis of the cylinder, and an
extraordinary wave when it lies in the plane formed by the vectors
$\bm k$ and $\bm a$ as it is shown in Fig. \ref{fig1} (a). In the
first case the differential cross section is written as
 \bea
\sigma_\perp=\frac{\pi k^2 R^4(\varepsilon
-1)^2}{2(\varepsilon+1)^2}\Biggl(2+2 \zeta \biggl(\text{Ci}(k L
(\zeta+1)-\text{Ci}(k L (1-\zeta)))~~~~~~~~~~~~~~\nonumber\\+\log
\left(\frac{2}{\zeta+1}-1\right)\biggr)-\frac{2 \sin (k L) \cos (k
L
\zeta)}{k L}~~~~~~~~~~~~~~~~~~~~~~~~~~~~~\nonumber\\
+\left(\frac{1+\zeta^2}{1-\zeta^2}\right)(z \sin (k L) \sin (k L
\zeta)+\cos (k L) \cos (k L \zeta)-1)\nonumber\\+(1+\zeta^2) k L
(\text{Si}(k L (\zeta+1))+\text{Si}(k L-k L \zeta))\Biggr),
\label{ext1}
\eea
where $\zeta=(\bm k\bm a)/k=\cos \phi$, $\mbox{Ci}$ is the cosine
integral function and  $\mbox{Si}$ is sine integral function.

 For
an extraordinary wave one has \bea \sigma_\parallel=\frac{\pi k^2
R^4(\varepsilon -1)^2}{2(\varepsilon+1)^2}\Biggl(\left((\epsilon
+1)^2-\zeta^2 (\epsilon  (\epsilon +2)+3)\right)
\biggl(\zeta\biggl(\text{Ci}(k L (1-\zeta))\nonumber\\-\text{Ci}(k
L (\zeta+1))+\log
\left(\frac{\zeta+1}{1-\zeta}\right)\biggr)-1\biggr)+\frac{\sin (k
L) \cos (k L \zeta)}{k L}\biggr)~~~~~~~~\nonumber\\+\left(\zeta^4
(\epsilon (\epsilon +2)+3)-2 \zeta^2 \epsilon  (\epsilon
+2)+(\epsilon
+1)^2\right)~~~~~~~~~~~~~~~~~~~~~~\nonumber\\\biggl(\frac{\zeta
\sin (k L) \sin (k L \zeta)+\cos (k L) \cos (k L
\zeta)-1}{1-\zeta^2}~~~~~~~~~~~~~~~\nonumber\\+\frac{k
L}{2}(\text{Si}(k L (\zeta+1))+\text{Si}(k
L(1-\zeta)))\biggr)\Biggr). \label{ext2} \eea

One could introduce the extinction cross sections per unit length
of the finite cylinder as $C_{\perp GRG}=\sigma_\perp/L$ and
$C_{\parallel GRG}=\sigma_\parallel/L$ in order to compare them
with that for the infinite one.

\section{Scattering by an infinite cylinder and the amplitude renormalization}

For  an infinite cylinder the extinction cross section is
expressed through the forward scattering matrix
\cite{hulst,boren}. Let us write a more general the "off shell"
expression for the forward scattering matrix \cite{cher}. It
arises when a wave propagates in the medium consisting of parallel
dielectric cylinders. The positions of the cylinders are suggested
to be uncorrelated. The dispersion equation \cite{cher} for the
effective wave vector $\bm k^\prime$  could be put into the form
\be k^{\prime 2}-k^2=4 i n_0 T(\bm k^\prime),
\label{eqras} \ee where $n_0$ is the concentration of the centers
of the cylinders in a plane perpendicular to the cylinder axis,
$k$ is a wave number of the an ectromagnetic wave in vacuum,
$T(\bm k^\prime)$ is the "of shell" forward scattering matrix. It
is considered that in the random medium there exists the mean
field, arising as a result of the multiple scattering of waves
\cite{lax} by the cylinders. In the Eq. \eqref{eqras} it is taken
into account that the wave falling on a cylinder in a random
medium differs from that in vacuum, thus, the "off shell" forward
scattering amplitude appears \cite{lax}.

\begin{figure}[th]
\hspace{-0. cm}
\includegraphics[width=10.4 cm]{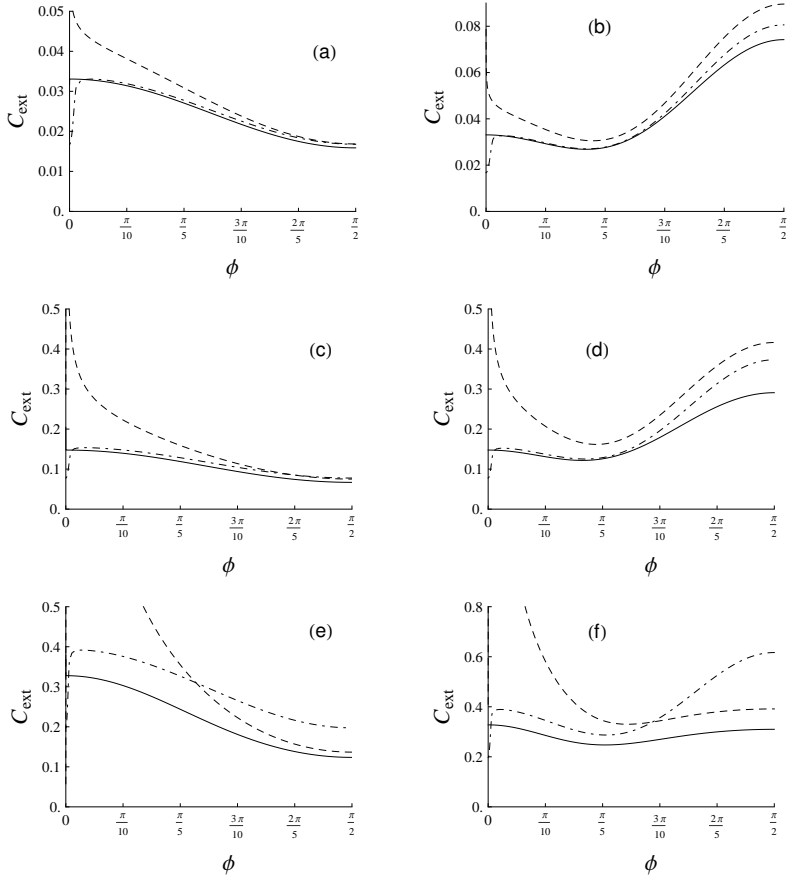}
\caption{\label{fig2} Extinction cross section per unit length of
the cylinder (a,b)-$\varepsilon=2.1$, $kR=0.3$,
(c,d)-$\varepsilon=2.1$, $kR=0.5$ ,(e,f)-$\varepsilon=1.5$,
$kR=1$. Figs. (a,c,e) and (b,d,f) correspond to the ordinary and
extraordinary waves respectively. Dashed-dotted line - GRG
approximation, dashed line - infinite cylinder without
renormalization, solid line - renormalized extinction by the
infinite cylinder. Aspect ratio $L/(2R)=10^4$ for GRG, $\phi$ is
the angle between wave number $\bm k$ and axis of the of the
cylinder. }
\end{figure}

In particular, if the concentration $n_0$ is low \be\bm k^\prime
\approx \bm k(1+2i n_0 T(\bm k)/k^2), \label{kp} \ee where $T(\bm
k)$ is the "on shell" value of the forward scattering matrix
$T(\bm k^\prime)$ when $\bm k^\prime$ is applied equal to $\bm k$.
From Eq. \eqref{kp} it follows $ \mbox{Im} \,k^\prime = C_{ inf}\,
n_0/2 $, where \be C_{ inf}=\frac{4}{k}Re[T(\bm k)] \label{cc} \ee
is the extinction cross section per unit length of the infinite
cylinder. According to \eqref{kp}, \eqref{cc} the intensity of the
wave decreases as the $I=I_{0}\exp\left({-C_{inf}\, n_0
\,l}\right)$ when the wave propagates in a random medium
consisting of parallel dielectric cylinders. Here $l$ is the path
length in the $\bm k$ direction.

\begin{figure}[th]
\hspace{-0. cm}
\includegraphics[width=9.5 cm]{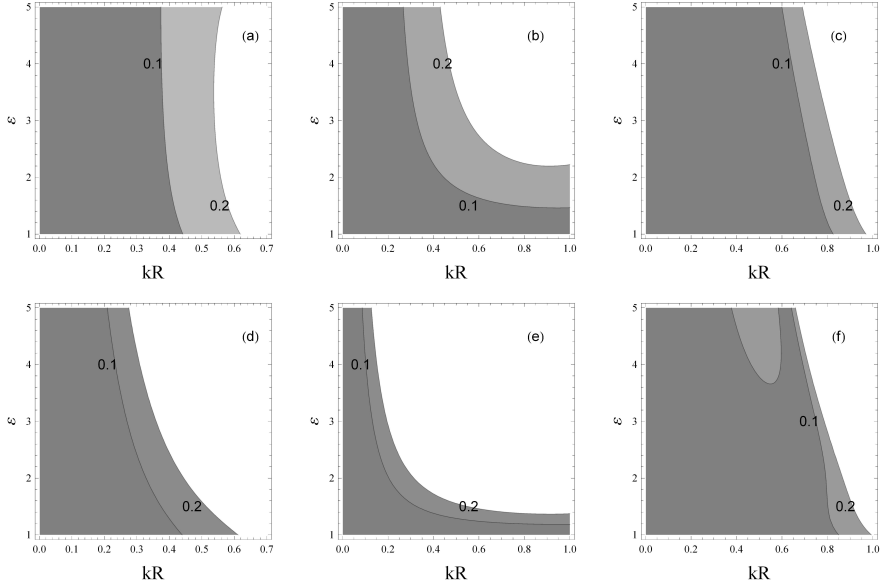}
\caption{\label{fig3} The ranges where the extinction errors of
the various approximations do not exceed 10 and 20 percents
relative to the renormalized extinction by infinite cylinder
(a),(d) -for GRG approximation, (b),(e) - for an infinite cylinder
without renormalization, (c),(f) - the expansion to the series
given by Eqns. \eqref{t1}, \eqref{t2}. Figs. (a),(b),(c)
correspond to an ordinary wave, (d),(e),(f) correspond to an
extraordinary wave. The aspect ratio is $L/(2R)=10^4$ for GRG. The
results is for the normal wave falling to the cylinder. }
\end{figure}

In the general case "off shell" expressions
 for the forward scattering matrix is given by \cite{cher}
\bea
T_{\perp}(\bm k^\prime)=\frac{\pi}{4
i}\sum_{n=0}^{\infty}\frac{\chi^-_n\frac{k_z^\prime}{k}\left(\delta^+_n\alpha_n+\frac{k_z^\prime}{k}\delta^-_n\beta_n\right)+\chi
^+_n \left(\delta_n^+\gamma_n+\frac{k_z^\prime}{k}\delta^-_n
\alpha_n\right)}{{\mathcal D}_n} \label{tpe}
   \eea
   for an ordinary wave  and
\bea
T_{\parallel}(\bm k^\prime)=\frac{\pi}{4
i}\sum_{n=0}^{\infty}\frac{\chi^+_n\frac{k_z^\prime}{k}\left(\delta^-_n\alpha_n+\frac{k_z^\prime}{k}\delta^+_n\beta_n\right)+\chi
^-_n \left(\delta_n^-\gamma_n+\frac{k_z^\prime}{k}\delta^+_n
\alpha_n\right)+\chi^0_n\delta^0_n\beta_n}{{\mathcal D}_n}
\label{tpr}
   \eea for an extraordinary wave. It is implied that
 $z$-axis is directed along the axis $\bm a$ of a cylinder, i.e.
 $k_z^\prime$ is the component of $\bm k^\prime$ parallel
to $\bm a$ and $\bm k_\perp^\prime$ is perpendicular to $\bm a$ as
it shown in Fig. \ref{fig1} (b).

Other  notations in the Eqs. \eqref{tpe}, \eqref{tpr} are
\bea
\alpha_n=n\frac{k_z^\prime}{k}\left(\frac{\lambda^2}{v^2}-1\right)H_n(v
R)J_n(\lambda R), ~v=\sqrt{k^2-k_z^{\prime
2}},~\lambda=\sqrt{\varepsilon k^2-k_z^{\prime 2}},\nonumber\\
\beta_n=\lambda  R \left(\frac{\lambda}{2 v}\left(H_{n-1}(v R)-H_{n+1}(v R)\right)J_n(\lambda R)-H_n(v R)J_n^\prime(\lambda R)
\right), \nonumber
\eea
\bea
 \gamma_n=\lambda R\left(\frac{\lambda}{2
v}(H_{n-1}(v R )-H_{n+1}(v R))J_n(\lambda R)-\varepsilon H_n(v
R)J_n^\prime(\lambda
R)\right),\nonumber\\
\delta^\pm_n=\frac{\lambda k^2(\varepsilon
-1)}{v(\lambda^2-k_\perp^{\prime 2})}\Biggl(k_\perp^\prime R
H_{n+1}(v R)J_{n+1}^\prime(k_\perp^\prime
R)~~~~~~~~~~~~~~~~~~~~~~~~~~~~~~~~\nonumber\\-\frac{v
R}{2}\left(H_n(v R)-H_{n+2}(v R)\right)J_{n+1}(k_\perp^\prime R)
\pm\biggl(k_\perp^\prime R H_{n-1}(v
R)J_{n-1}^\prime(k_\perp^\prime R)\nonumber\\-\frac{v
R}{2}\left(H_{n-2}(v R)-H_n(v R)\right)J_{n-1}(k_\perp^\prime
R)\biggr)\Biggr),\nonumber
\eea
\bea
\delta^0_n=\frac{2(\varepsilon-1)\lambda^2}{\lambda^2-k_\perp^{\prime
2}}\biggl(k_\perp^\prime R H_n(v R)J_n^\prime(k_\perp^\prime
R)~~~~~~~~~~~~~~~~~~~~~~~~~~~~~~~~~~~~~~~\nonumber\\-\frac{v
R}{2}\left(H_{n-1}(v R)-H_{n+1}(v R)\right)J_n(k_\perp^\prime
R)\biggr),\nonumber \eea
\be
 \chi_n^\pm = \left\{
{{\begin{array}{*{20}c}
 {\left(k_\perp^\prime R J_{n+1}(\lambda R)J_{n+1}^\prime(k_\perp^\prime R)-\lambda R J_{n+1}(k_\perp^\prime R)J_{n+1}^\prime(\lambda
 R)\right)}
  {} \hfill \\
 \\{\pm \left(k_\perp^\prime R J_{n-1}(\lambda R)J_{n-1}^\prime(k_\perp^\prime R)-\lambda R J_{n-1}(k_\perp^\prime R)J_{n-1}^\prime(\lambda R)\right), n\ne 0,} \hfill \\
 {} \hfill \\
 {(1\pm1)/2\left(k_\perp^\prime R J_{1}(\lambda R)J_{1}^\prime(k_\perp^\prime R)-\lambda R J_{1}(k_\perp^\prime R)J_{1}^\prime(\lambda R)\right) ,n =
0,} \hfill \\
\end{array}} } \right.\nonumber
\ee
\bea
\chi_n^0 = \left\{ {{\begin{array}{*{20}c}
 {2\left(k_\perp^\prime R J_{n}(\lambda R)J_{n}^\prime(k_\perp^\prime R)-\lambda R J_{n}(k_\perp^\prime R)J_{n}^\prime(\lambda
 R)\right),n \ne 0,} \hfill \\
 {} \hfill \\
 {\left(k_\perp^\prime R J_{0}(\lambda R)J_{0}^\prime(k_\perp^\prime R)-\lambda R J_{0}(k_\perp^\prime R)J_{0}^\prime(\lambda
 R)\right) ,n =
0,} \hfill \\
\end{array}} } \right.\nonumber\\
{\mathcal
D}_n=\alpha_n^2-\beta_n\gamma_n.~~~~~~~~~~~~~~~~~~~~~~~~~~~~~~~~~~\nonumber
\eea

For the "on shell" scattering, when $k^\prime=\sqrt{k_z^{\prime
2}+\bm k_\perp^{\prime 2}}=k$, the amplitudes $T_\perp(\bm
k^\prime)$ and $T_\parallel(\bm k^\prime)$ are reduced to the
amplitudes $T_\perp(\bm k)$ and $T_\parallel(\bm k)$  describing
forward light scattering by a single infinite cylinder. They
presented, for instance, in Refs. \cite{hulst,boren}. Let us
calculate numerically the extinction cross section divided by the
length of the cylinder in the GRG approximation (dashed-doted
lines in Fig. \ref{fig2}) and compare it with that for infinite
cylinder (dashed lines in Fig. \ref{fig2}). As one can see from
Fig. \ref{fig2} the cross sections are close near the
perpendicular incidence and differ substantially when the
direction of the incident wave is close to the axis of the
cylinder. There exists the narrow gap, in the GRG approximation
related to the finiteness of the cylinder, however, even outside
of this gap the curves remains far one from another. Moreover, the
extinction cross section for the infinite cylinder diverges
logarithmically  in the vicinity of small $k_\perp$. As for the
GRG approximation, it does not contain logarithmic terms
asymptotically at $L\rightarrow \infty$ and $k_\perp>>1/L$. This
observation leads to an idea of removing all the logarithmic terms
from the forward scattering amplitude by the infinite cylinder.
Fortunately it is easy to do because the logarithmic terms
originate from the Hankel functions. According to \cite{grad} the
Hankel function of the first kind could be represented in the form
\be H_n(z)=J_n(z)+iY_n(z), \ee where $J_n(z)$, $Y_n(z)$ are the
Bessel functions of the first and the second kind respectively.
Regular part of the function $Y_n(z)$ could be extracted with the
help of  the formula \cite{grad} \bea \pi Y_n(z)=2
J_n(z)\left(\ln\frac{z}{2}+\gamma\right)-\sum_{k=0}^{n-1}\frac{(n-k-1)!}{k!}\left(\frac{z}{2}\right)^{2k-n}~~~~~~~\nonumber\\
-\left(\frac{z}{2}\right)^n\frac{1}{n!}\sum_{k=1}^{n}\frac{1}{k}-\sum_{k=1}^\infty
\frac{(-1)^k}{k!(k+n)!}\left(\frac{z}{2}\right)^{n+2k}\left(\sum_{m=1}^{n+k}\frac{1}{m}+\sum_{m=1}^k\frac{1}{m}\right),
\eea
where $\gamma$ is Euler constant. As a result, renormalized Hankel
function without logarithmic terms is
\be
H_n^{ren}(z)=J_n(z)+i\left(Y_n(z)-\frac{2}{\pi}J_n(z)\ln\frac{z}{2}\right).
\label{hren}
\ee

Let us calculate the "on shell" forward scattering matrix and
corresponding extinction cross section using  formulas
\eqref{tpe}, \eqref{tpr} and renormalized Hankel function
\eqref{hren}. It should be emphasized that before the
renormalization the amplitudes \eqref{tpe}, \eqref{tpr} have been
written in the form containing Hankel functions but not their
derivatives.

 As it is shown in
Fig.\ref{fig2} the coincidence with the extinction in the GRG
approximation and that given by the infinite cylinder after
renormalization \eqref{hren} is rather well in a range where GRG
have to be valid. For an extraordinary wave the coincidence is not
so well, but probably, it is related with the applicability of the
GRG approximation.

Let us put the forward scattering amplitudes \eqref{tpe},
\eqref{tpr} "on shell" and expand them into the series  up to the
terms of the six order in $R$. Representing the resulting
amplitudes as $T_{ser}=T_{ser}^\prime+iT_{ ser}^{\prime\prime}$ we
come to \bea T_{\perp ser}^\prime=\frac{\pi ^2 (k R)^4 (\epsilon
-1)^2}{32 (\epsilon +1)^3} \Biggl((k R)^2 \biggl(2 \zeta ^2
\left(\epsilon ^2-1\right)+\zeta ^4 (\epsilon +3)\nonumber\\-4
\gamma \left(\zeta ^2+1\right)^2 (\epsilon -1)-\epsilon
-3\biggr)+4 \left(\zeta ^2+1\right)
   (\epsilon +1)\Biggr),
   \label{t1}
\eea
\bea
T_{\perp ser}^{\prime\prime}=-\frac{\pi  (k R)^2 (\epsilon -1)
}{384 (\epsilon +1)^3}\Biggl((k R)^4 (\epsilon -1) \biggl(-\zeta
^2+24 \gamma \Bigl(-2 \zeta ^2 \left(\epsilon
^2-1\right)\nonumber\\-\zeta ^4 (\epsilon +3)+\epsilon
+3\Bigr)+\zeta ^4 (23\epsilon-3 \epsilon^2
   +20)+\zeta ^2 \epsilon  (  2 \epsilon^2 -\epsilon-16)~~~~~~~~~~\nonumber\\-12 \pi ^2 \left(\zeta ^2+1\right)^2 (\epsilon -1)
   +48 \gamma ^2 \left(\zeta ^2+1\right)^2 (\epsilon -1)+\epsilon
   (2 \epsilon^2
   +8\epsilon-11)-23\biggr)\nonumber\\+12 (k R)^2 \left(\epsilon ^2-1\right) \biggl(\zeta ^2 (\epsilon -8 \gamma -1)+\epsilon -8 \gamma +3\biggr)
   +192 (\epsilon +1)^2\Biggr).
   \label{t2}
\eea

\begin{figure}[th]
\hspace{-0. cm}
\includegraphics[width=11.5 cm]{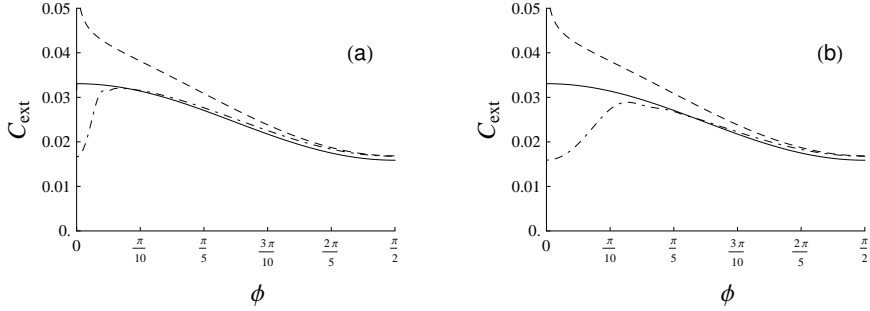}
\caption{\label{fig4} The same as in Fig. \ref{fig2}, (a) but for
the  aspect ratio used in GRG $L/(2R)=10^3$ - (a), $L/(2R)=10^2$ -
(b).}
\end{figure}

For an extraordinary wave one have \bea T_{\parallel
ser}^\prime=\frac{\pi ^2 (k R)^4 (\epsilon -1)^2}{64 (\epsilon
+1)^3} \Biggl((k R)^2 \biggl(4 \gamma \Bigl(\zeta ^6+7 \zeta
^4-\zeta ^2+\left(\zeta ^2-1\right)^3 \epsilon
^4~~~~~~~\nonumber\\+2 \left(\zeta ^2-1\right)^3 \epsilon ^3+2
\left(-2 \zeta ^6+\zeta ^4-4 \zeta
   ^2+1\right) \epsilon +1\Bigr)\nonumber~~~~~~\\+\zeta ^6 (3 \epsilon ^3+9 \epsilon ^2+15 \epsilon +13)
   +\zeta ^4 (\epsilon +1) (\epsilon ^3-7 \epsilon ^2-17 \epsilon -17)\nonumber\\
   +\zeta ^2 (3-2 \epsilon ^4+3 \epsilon ^3+25 \epsilon ^2+27 \epsilon)+(\epsilon -3) (\epsilon +1)^3\biggr)~~~
   ~~~~~\nonumber\\+4 (\epsilon +1) \Bigl(\zeta ^4
   (\epsilon^2+2\epsilon +3)-2 \zeta ^2 \epsilon  (\epsilon +2)+(\epsilon
   +1)^2\Bigr)\Biggr),
   \label{t3}
  \eea
  \bea
T_{\parallel ser}^{\prime\prime}=\frac{\pi  (k R)^2 (\epsilon
-1)}{384 (\epsilon +1)^3} \Biggl((k R)^4 (\epsilon -1)
\biggl(\zeta ^4 \Bigl(6 \pi ^2 \left(\epsilon  \left(3 \epsilon
^3+ 6\epsilon^2-2\right)-7\right)~~~~~~\nonumber\\+\epsilon (38
\epsilon^2 +101\epsilon+72)-3\Bigr)+\zeta
   ^6 \Bigl(-\bigl(6 \pi ^2 \left(\epsilon ^4+2 \epsilon ^3-4 \epsilon
   +1\right)
   \nonumber\\+\epsilon  (12\epsilon^2 +31\epsilon+43)+18\bigr)\Bigr)+\zeta ^2 \Bigl(6 \pi ^2 \left(-3 \epsilon ^4-6 \epsilon ^3
   +8 \epsilon
   +1\right)~~~\nonumber\\+\epsilon  (2 \epsilon  (\epsilon^2 -17\epsilon -49)-53)+15\Bigr)+24 \gamma ^2 \Bigl(\zeta ^6+7 \zeta ^4-\zeta ^2
   +\left(\zeta ^2-1\right)^3 \epsilon ^4\nonumber\\+2 \left(\zeta ^2-1\right)^3 \epsilon ^3+2
   \left(-2 \zeta ^6+\zeta ^4-4 \zeta ^2+1\right) \epsilon +1\Bigr)+12 \gamma  \Bigl(\zeta ^6
   (3 \epsilon^3+9\epsilon^2\nonumber\\+15\epsilon+13)+\zeta ^4 (\epsilon +1) (\epsilon  (\epsilon^2 -7\epsilon
   -17)-17)+\zeta ^2 (\epsilon  ((3\epsilon^2-2 \epsilon^3  +25\epsilon)\nonumber\\+27)+3)+(\epsilon -3) (\epsilon +1)^3\Bigr)
   +2 \left(3 \pi ^2 (\epsilon -1)-\epsilon +5\right) (\epsilon +1)^3\biggr)\nonumber\\+12 (k R)^2
   \left(\epsilon ^2-1\right) \biggl(\zeta ^4 (3 \epsilon +4 \gamma  (\epsilon  (\epsilon +2)+3)+5)
   +\zeta ^2 (\epsilon  (\epsilon -8 \gamma  (\epsilon +2)-3)\nonumber\\-6)+(4 \gamma -1) (\epsilon +1)^2\biggr)+96 (\epsilon
   +1)^2 \biggl(\zeta ^2 (\epsilon -1)-\epsilon -1\biggr)\Biggr).
   \label{t4}
\eea The equations \eqref{t1},\eqref{t2},\eqref{t3},\eqref{t4} do
not contain any logarithmic terms because they were  removed by
the renormalization of the Hankel function. In comparison with the
extinction, given by the GRG approximatioin,  the equations
\eqref{t1},\eqref{t3} contain the terms of the six order in $R$
and, thus, they have the wider applicability for the shorter
wavelength.

In the general case, if to consider falling of the electromagnetic
wave on the layer of cylinders not only the extinction cross
section is needed but the effective refractive index of a layer.
It can be found from the dispersion equation for the wave vector
\eqref{eqras} using renormalized Hankel functions.

\section{Discussion and conclusion}

The expressions \eqref{t1}, \eqref{t2}, \eqref{t3}, \eqref{t4} are
of the six order in $R$. The expressions \eqref{ext1},
\eqref{ext2} are of the fourth order in $R$. Calculating the
asymptotic of the $\sigma/L$  given by \eqref{ext1}, \eqref{ext2}
in the limit $L\rightarrow \infty$, $k_\perp>>1/L$ we have found
that the extinctions per unit length of the finite cylinder
coincides exactly with the corresponding $R^4$ terms contained in
the Eq. \eqref{t1}, \eqref{t3}.

The fascinating hypothesis arises that in the limit $L\rightarrow
\infty$, $k_\perp>>1/L$ the extinction cross section as a result
of the exact solution of the three dimensional scattering problem
coincides exactly, i.e. in the all orders in $R$, with the
extinction by the infinite cylinder obtained with the renormalized
Hankel functions. From  physical point of view it means that with
the help of renormalization of the Hankel functions one is able
 to describe correctly the extinction by the long but finite dielectric
cylinder in all orders in $kR$ and in all the range of the
incidence angles excluding the narrow gap near $\phi=0$.

We have no possibility to check this general  hypothesis
analytically, thus, we have  presented numerical results showing
the ranges, where the relative errors compared to $GRG$ and the
infinite cylinder without renormalization approximations do not
exceed 10 per and 20 percents. The normal incidence of a wave to
the cylinder is considered in order to check  also the earlier
hypothesis of van der Hulst \cite{hulst,wang} that at a normal
wave incidence an extinction by the finite cylinder is equal to
that by the infinite cylinder (without renormalization). As one
could see the hypothesis of van der Hulst is approximate, because
the logarithmic terms spoil the coincidence even at the normal
incidence, when the dielectric constant and $kR$ become large. At
the same time from the Fig. \ref{fig3} we see that the GRG
approximation and
 the approximate formulas \eqref{t1},\eqref{t3}
work at a  relatively wide range.

As it was mentioned the gap for the small angles of incidence
relative the axis of the finite cylinder could not principally be
described in a frame of scattering by infinite cylinder. From the
Fig. \ref{fig4} one could see this gap broadening with the rising
of the aspect ratio. However, at low aspect ratio the numerical
methods can be easily applied.

We may conclude that the renormalized Hankel functions should be
used to describe the extinction as by a single cylinder so in the
dispersion equations for the effective wave number \eqref{eqras}.
The dispersion equation \eqref{eqras} could be applied
straightforwardly to the mediums where the correlation in the
placement of the scatterers are absent, for instance, porous
silicon \cite{pap}. However, for instance, for porous aluminium
oxide \cite{gap} the correlation of the scatterers should be taken
into account \cite{cher2}  with the help of radial distribution
function \cite{cryst}.

\end{document}